\documentclass{revtex4-1}
\usepackage[latin9]{inputenc}
\usepackage[active]{srcltx}
\usepackage{amssymb}
\usepackage{graphicx}

\makeatletter

\@ifundefined{textcolor}{}
{%
 \definecolor{BLACK}{gray}{0}
 \definecolor{WHITE}{gray}{1}
 \definecolor{RED}{rgb}{1,0,0}
 \definecolor{GREEN}{rgb}{0,1,0}
 \definecolor{BLUE}{rgb}{0,0,1}
 \definecolor{CYAN}{cmyk}{1,0,0,0}
 \definecolor{MAGENTA}{cmyk}{0,1,0,0}
 \definecolor{YELLOW}{cmyk}{0,0,1,0}
}

\makeatother

\begin{document}

\preprint{This line only printed with preprint option}

\title{Physics of quantum measurement and its interdisciplinary applications}

\author{Masahiro Morikawa}

\email{hiro@phys.ocha.ac.jp}

\affiliation{Department of Physics, Ochanomizu University, Tokyo 112-0012 JAPAN}

\author{Akika Nakamichi}

\email{nakamichi@cc.kyoto-su.ac.jp}

\affiliation{Koyama Observatory, Kyoto-Sangyo University, Kyoto 603-8555 JAPAN}
\begin{abstract}
Quantum dynamics of the collective mode and individual particles on
a ring is studied as the simplest model of projective quantum measurement.
In this model, the collective mode measures an individual single quantum
system. The heart of the model is the wide separation of time scales
which yields the distinction of classical and quantum degrees of freedom
beyond the standard Gross-Pitaevskii equation. In some restricted
cases we derive the Born probability rule. This model is the quantum
mechanics version of the effective action method in quantum field
theory, which describes the origin of the primordial density fluctuation
as classical variables. It turns out that the classical version of
this same model successfully describes the dynamics of geomagnetic
variation including the polarity flips over 160 million years. The
essence of this description is again the coexistence of the wide separated
time scales. 
\end{abstract}
\maketitle

\section{introduction}

Quantum mechanics is an excellent theory to describe the Universe
and have never failed in laboratory experiments. . However in more
fundamental level when an individual wave function is considered,
especially in relation with delicate measurement processes, the operational
description of quantum mechanics is not sufficient and sometime leads
us to confusion. A typical situation is the description of the entire
Universe by a single wave function. Although the Wheeler-DeWitt equation
with any favorite boundary condition formally yields a definite form
of the wave function, we have no justified treatment of the wave function.
We cannot repeat the evolution process of the Universe nor take any
interpretation based on the frequency distribution. More serious problem
is the calculation of the primordial density fluctuations in the early
Universe. It may be formally possible to calculate the two point correlation
function in quantum mechanics for a k-mode inflaton $\left\langle \hat{\phi}\left(x\right)\hat{\phi}\left(y\right)\right\rangle \left|_{k}\right.$
however we have no justification how the spatially inhomogeneity and
the statistical power spectrum are related with $\left\langle \hat{\phi}\left(x\right)\hat{\phi}\left(y\right)\right\rangle \left|_{k}\right.$.

All the above problems seems to stem from the hybrid structure of
quantum mechanics: deterministic time evolution described by Schr�dinger
equation and the stochastic measurement process where the probability
enters. The latter process cannot be described only by the former
since the latter is not deterministic nor linear. Therefore an extra
projection postulate is introduced in quantum mechanics. There have
been variety of study on the measurement process so far. We would
like to examine a natural physical process based on the collective
motion without changing the present formalism of quantum mechanics
at all. The measurement process should be physical and it is natural
to describe it by the Schr�dinger equation at least in the fundamental
level. In this paper, we focus on the interaction between the collective
mode and the individual degrees of freedom in a simple model.

Collective motion is a common phenomena in various complex systems
of many degrees of freedom\cite{key-1}\cite{key-2}. It means the
formation of a localized condensation of the constituent degrees of
freedom in the system. This condensation is generally dynamical and
appears as a macroscopic degrees of freedom, which we call the order
variable. The evolution equation of the order variable is generally
different from the one for the individual degrees of freedom and can
be non-linear even if the latter is linear. Moreover the appearance
of the order variable is often characterized by the separation of
its time scale from that of individual degrees of freedom. Actually
we successfully analyzed the long-term behavior of geomagnetism based
on this line of thought\cite{key-13}. These two types of degrees
of freedom interact with each other and one determines the other self
consistently. This self-consistent description brings non-trivial
dynamics which was not in the original system. The same is true for
quantum theory.

There have been many arguments on the basic part of quantum mechanics\cite{key-3}\cite{key-4}.
We simply focus on the problem how the detector works based on the
Schr�dinger equation in this paper. An important point we would like
to emphasize is that the ordinary decoherence mechanism is not at
all sufficient to describe the resultant system state after one-shot
measurement. Our goal will be to deduce the result consistent with
the standard measurement postulate in quantum mechanics and clarify
the physical process actually happening in the detector for quantum
measurement. Although a general argument for this purpose utilizing
quantum field theory and the generalized effective action has been
explored before\cite{key-10}\cite{key-11}%
\footnote{A classical Gross-Pitaevskii equation with fluctuation and dissipation
is derived and used to describe the measurement appararus. However
the coupling of this classical field and the quantum measured system
was not fully analized. %
}, present paper is an attempt to study fully within the theory of
quantum mechanics. We would like to emphasize how the time scale separation
emerges and characterize the detector degrees of freedom in the quantum
mechanical argument without resorting to the infinite degrees of freedom
of quantum field theory from the beginning.

The next section 2 describes a simple model of measuring apparatus
which is described by Schr�dinger equation with the self-consistent
approximation. We emphasize the distinction between the quantum and
classical like behaviors in this system. Section 3 describes the response
of the detector by the interaction with the quantum system which is
measured, as well as the back reaction of the detector to the system.
Section 4 describes the detail of the quantum measurement for the
system of general superposition. The last section 5 is devoted to
the summary of the work and to the setup of the future studies.

\section{quantum ring model}

\begin{figure}
\includegraphics[height=5cm]{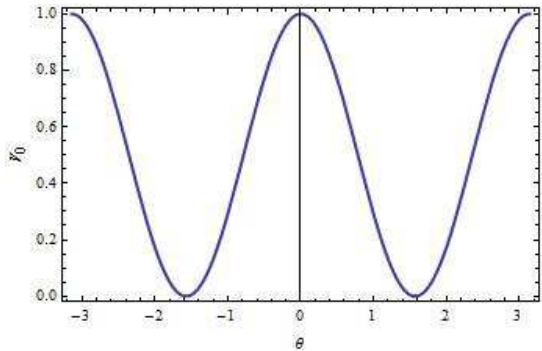}\caption{The graph shows the potential $V_{0}\left(\theta\right)=cos(\theta)^{2}$.
It has two stable points at $\theta=\pm\pi/2$. The initial wave functions
are set around the unstable point very near to $\theta=0$. The wave
function eventually falls down partially into left and partially into
right.\label{fig:1}}
\end{figure}

We consider the wave function for large $N$ particles moving on a
ring of unit radius. No method is known so far to solve this system
exactly in general. Furthermore, few exactly-solvable models are exceptional
and will not represent general detector apparatus. On the other hand,
we have to introduce an environment or its equivalent, after all,
to represent the actual detector which decoheres and is non-deterministic.
This makes the exactly solvable model unnecessary and allow flexible
approximations to solve the model. We use a single tensor product
of the wave functions for the individual particles $\psi_{i}(t,\theta_{i}),\; i=1,2,...,N$
$(-\pi\leq\theta_{i}<\pi)$, similar to the time-dependent Hartree-Fock
approximation. The particles are in the common potential $V_{0}\left(\theta_{i}\right)=cos(\theta_{i})^{2}$
(\ref{fig:1}) and interact with each other with the attractive force%
\footnote{The attractive interaction generally represents any mechanism which
cause condensation. %
}.

The potential$V_{0}\left(\theta\right)$has two stable points at $\theta=\pm\pi/2$.
We will set the initial location of many wave functions very near
to the unstable point $\theta=0$ with small finite variance$\sigma$.
The attractive interaction can be represented by an extra potential
field, which is yielded by all the other particles, as 
\begin{equation}
V_{HF}\left(\theta_{i}\right)\equiv\lambda\left(\varphi(t,\theta_{i})^{2}-\frac{1}{N}\left|\psi_{i}(t,\theta_{i})\right|^{2}\right)\label{eq:1}
\end{equation}
for the $i$-th particle ($\lambda<0$), where 
\begin{equation}
\varphi(t,\theta)^{2}\equiv\frac{1}{N}\sum_{k=1}^{N}\left|\psi_{k}(t,\theta)\right|^{2}\label{eq:2}
\end{equation}
is defined to be the order variable.

Then the equation of motion for the wave function of individual $i$-th
particle $\psi_{i}(t,\theta_{i})$ becomes 
\begin{equation}
i\hbar\frac{\partial\psi_{i}(t,\theta_{i})}{\partial t}=-\frac{\hbar^{2}}{2m}\frac{\partial^{2}}{\partial\theta_{i}^{2}}\psi(t,\theta_{i})+V_{0}\left(\theta_{i}\right)\psi(t,\theta_{i})+V_{HF}\left(\theta_{i}\right)\psi_{i}(t,\theta_{i}),\label{eq:3}
\end{equation}
where the exchange Fock potential is not necessary in our boson case.
The order variable $\varphi(t,\theta)$ and the individual particle
$\psi_{i}(t,\theta_{i})$ cooperatively determine their evolution.
This self-contained structure makes the effective non-linear feature
in the set of equations while individual wave function $\psi_{i}(t,\theta_{i})$
still maintains its linearity since the coupling factor Eq.(\ref{eq:1})
to $\psi_{i}(t,\theta_{i})$ in Eq.(\ref{eq:3}) does not contain
itself $\psi_{i}(t,\theta_{i})$.

If the attractive force ($\lambda<0$) is strong enough and dominates
the total energy, then the mean field$\varphi(t,\theta)^{2}$ tends
to be localized in $\theta-$space irrespective of the common potential
$V_{0}\left(\theta_{i}\right)$. This localized collective mode evolves
keeping its locality. Then we naturally call this variable as order
variable, which spontaneously violates the spatial ($\theta$) translational
invariance.

Since we set the initial location of many wave functions very near
to the unstable point $\theta=0$ with small finite variance$\sigma$,
the condition for this localized collective mode to appear may become
\begin{equation}
V_{0}<V_{HF}.\label{eq:4}
\end{equation}

Furthermore the quantum tunneling probability of the localized order
variable $\varphi(t,\theta)^{2}$ to penetrate the potential $V_{0}\left(\theta\right)$
can be extremely small compared with that of a single particle $\psi_{i}(t,\theta_{i})$,
\begin{equation}
\exp\left[-\frac{\sqrt{Nm\triangle E/2}a}{\hbar}\right]=e^{-\left(\sqrt{N}-1\right)\sqrt{m\triangle E/2}\left(a/\hbar\right)}\exp\left[-\frac{\sqrt{m\triangle E/2}a}{\hbar}\right],\label{eq:5}
\end{equation}
for large $N$, where $\triangle E,\, a$ respectively represent the
typical energy per particle and the typical size of the potential
$V_{0}$. This fact yields the separation of time scales in the whole
system, as claimed in the introduction. If we choose the measurement
time scale $T_{measurement}$ as 
\begin{equation}
\left[\exp\left(-\frac{\sqrt{Nm\triangle E/2a}}{\hbar}\right)\right]^{-1}\gg T_{measurement}>\left[\exp\left(-\frac{\sqrt{m\triangle E/2a}}{\hbar}\right)\right]^{-1},\label{eq:6}
\end{equation}
then the order variable $\varphi(t,\theta)^{2}$ behaves as a single
classical degrees of freedom while the individual particles evolve
as quantum mechanically. Thus the whole system has quantum and classical
variables simultaneously and they are consistently described by the
set of equations Eqs.(\ref{eq:1}-\ref{eq:3}) provided the limited
time scale $T_{measurement}$ is considered.

\section{detector readout - no trigger}

\begin{figure}
\includegraphics[width=10cm]{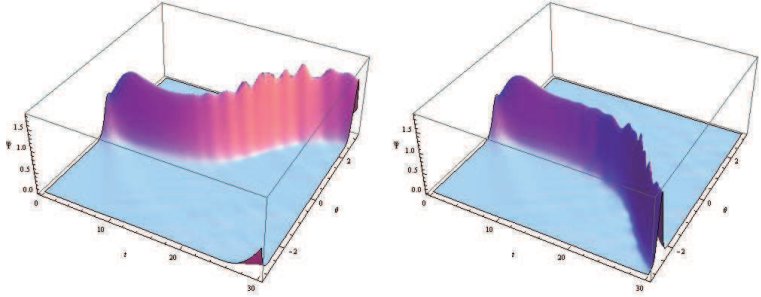}\caption{Time evolution of the order variable $\varphi(t,\theta)^{2}$ with
$N=100$. We set the initial Gaussian wave functions of the particles
at $\theta\approx0$ with their peaks have small dispersion $\sigma$.
The attractive force between the particles makes the whole wave functions
to form a coherent cluster which evolves as an almost single Gaussian
mode. It eventually falls down the potential valley toward either
to the positive (left)) or negative (right)) side. The destination
is fully random reflecting the initial random distribution of the
wave functions. However this system can be triggered by the system
particle $i=0$, which is set as Eq.(\ref{eq:9-1}). The accuracy
of the numerical calculations is indicated by the energy conservation
within about 0.1\%. \label{fig:2}}
\end{figure}

We now demonstrate the time evolution of the wave functions based
the Eqs.(\ref{eq:1}-\ref{eq:3}). We are especially interested in
the correlation between the order variable $\varphi(t,\theta)^{2}$
and individual variables $\psi_{i}(t,\theta_{i}),\; i=1,2,...,N$.
We prepare the initial wave functions as 
\begin{equation}
\psi_{i}(t=0,\theta)=\frac{1}{\sqrt{2\pi}s}\exp\left(-\frac{\left(\theta-\xi_{i}\right)^{2}}{2s^{2}}\right)e^{i\xi_{i}'\theta}\label{eq:7}
\end{equation}
for $i=1,2,...,N$ with $N=100$, and $\xi_{i}$, $\xi_{i}'$ are
random variables which obey Gaussian distributions with the center
naught and the dispersion $\sigma=10^{-1}$. Other parameters are
$\hbar=0.02,\: m=1,\: s^{2}=10^{-1}$. Thus all the wave packets are
prepared very near around $\theta\approx0$. A typical time evolution
of $\varphi(t,\theta)^{2}$ is shown in Fig.\ref{fig:2}.

Individual particles are very unstable around the position $\theta\approx0$,
even the peak is exactly located at the top of the potential $\theta=0$
since the wave packets Eq.(\ref{eq:7}) bifurcate toward the both
valleys of the potential. However actually in our model, particle
are attractive with each other and they tend to be gather to form
a localized cluster which evolves coherently. This is the behavior
shown in Fig.\ref{fig:2}.

The localized cluster evolves into either sides of the valley, but
not both, for each shot of calculations. This probabilistic feature
enters in our model at the initial preparation of the wave functions;
the peak locations and the phases of them are chosen randomly. The
condition of this collective evolution to occur is given by Eq.(\ref{eq:4}).

Actually in our case of Eq.(\ref{eq:7}), this condition becomes 
\begin{equation}
\frac{\sigma^{2}}{2m}+\left|\sigma V_{0}'(\sigma)\right|<\frac{\left|\lambda\right|}{2},
\end{equation}
which is satisfied in the calculations shown in Fig.\ref{fig:2}.

\section{dynamics of measurement process}

\subsection{single trigger}

We now apply this model to the dynamics of quantum measurement. We
introduce an extra particle, $i=0$, into the above model as the system
which is measured and the original particles, $i=1,2,...,N$, are
the apparatus. We only measure whether the system particle is located
in the positive region or negative region. The evolution equations
Eqs.(\ref{eq:1}-\ref{eq:3}) hold as before, except that the summation
runs on $i=0,1,2,...,N$ in Eq.(\ref{eq:2}). Original Eq.(\ref{eq:2})
however, still serves as the definition of the order variable for
the apparatus. In summary, the system which is measured is the particle
$i=0$ and the apparatus is the particles $i=1,2,...,N$. The meter
readout is given by Eq.(\ref{eq:2}). This is a model of non-shot
measurement of a quantum state.

\begin{figure}
\includegraphics[width=10cm]{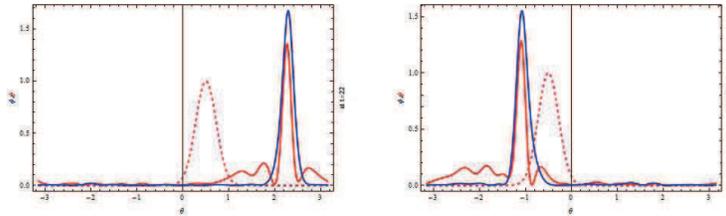}\caption{Snapshots of the system wave function squared $\left|\psi_{0}(t,\theta)\right|^{2}$(solid
red line) and the order variable $\left|\varphi(t,\theta)\right|^{2}$(solid
blue line) at time $t=24$. The initial distribution $\left|\psi_{0}(t,\theta)\right|^{2}$(red
broken line) of the system wave function is also shown. The system
wave function $\psi_{0}(t,\theta)$ initially has Gaussian distribution
with its peak at $\theta=+0.5$ or $\theta=-0.5$. The apparatus system
is the same as Fig.\ref{fig:2}. Initially $\psi_{0}(t,\theta)$ triggers
$\varphi(t,\theta)^{2}$ and this $\varphi(t,\theta)^{2}$ attracts
$\psi_{0}(t,\theta)$ to make its peak follow the order variable $\varphi(t,\theta)^{2}$.
However, reflecting the fact that the whole system is conservative
and the number of particles $N=100$ being too small, the peak of
the wave function $\psi_{0}(t,\theta)$ is not sharp and the whole
wave function is dispersed in $\theta$ space.\label{fig:3}}
\end{figure}

We consider in this paper the following one-parameter series of wave
functions for the initial state of the system wave function 
\begin{equation}
\psi_{0}(t,\theta_{0})=\sin\left(\alpha\right)\exp\left(-\frac{\left(\theta+\frac{1}{2}\right)^{2}}{2s^{2}}\right)+\cos\left(\alpha\right)\exp\left(-\frac{\left(\theta-\frac{1}{2}\right)^{2}}{2s^{2}}\right).\label{eq:9-1}
\end{equation}
We first consider the case of a single trigger $\alpha=0$ and $\alpha=\pi/2$.
For the trigger $\alpha=0$, for example, the state wave function
locates near the right bottom of the potential valley. This equation
is solved simultaneously with Eqs.(\ref{eq:1}-\ref{eq:3}). The system
$i=0$ triggers the apparatus particles $i=1,2,...,N$, and attracts
them to the right valley. The system wave function itself tends to
move with the apparatus particles. Then all the particles fall down
toward the right valley without dispersing due to their attractive
force. Therefore the system settles down to the positive position
with the meter readout Eq.(\ref{eq:2}) in the positive side. (Fig.\ref{fig:3}
left)

If the system wave function is set near the left bottom of the potential
(\textit{i.e.} $\alpha=\pi/2$), then all the wave functions fall
down toward left. In this case, the system settles down to the negative
position with the meter readout Eq.(\ref{eq:2}) in the negative side
(Fig.\ref{fig:3} right). This dynamical process thus successfully
creates the correlation necessary for the quantum measurement process.

The consistency condition between the system wave function and the
order variable with respect to the right-or-left positions reduces
to the condition that the initial system wave function could actually
put sufficient effect for the order variable to roll down toward the
direction of the wave function. This condition may become 
\begin{equation}
\frac{\sigma}{\sqrt{N}}V'\left(\frac{\sigma}{\sqrt{N}}\right)<\lambda.
\end{equation}
In the present case in Fig.\ref{fig:3}, this inequality is satisfied.

The consistency of the locations of system wave function $\psi_{0}(t,\theta)$
and the meter reading $\varphi(t,\theta)^{2}$ is shown in Fig. \ref{fig:3}.
The system $\psi_{0}(t,\theta)$ is set on the right of the potential
eventually settle down toward on the same right side of the potential
consistently with the meter readout $\varphi(t,\theta)^{2}$.

\subsection{even trigger$\alpha=\pi/4$ : }

\begin{figure}
\includegraphics[width=10cm]{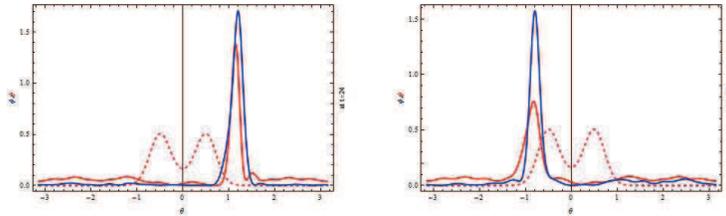}\caption{Same as Fig.\ref{fig:3}, but the system wave function $\psi_{0}(t,\theta)$
initially has evenly superposed two Gaussian distributions with peaks
at $\theta=+1/2$ and $\theta=-1/2$. \label{fig:4}}
\end{figure}

The above faithfulness of the order variable to the final state of
the system wave function appears also in the case $\alpha=\pi/4$.
Fig.\ref{fig:4}.

The important observation is that the double-peak wave function reduces
to a single-peak form, consistent with the order variable. The condition
for this reduction to take place is that the time scale of the measurement
$T_{measurement}$ is sufficiently longer than the typical time scale
for the system wave function to tunnel the potential. This condition
yields the right side of Eq.(\ref{eq:6}): 
\begin{equation}
T_{measurement}>\left[\exp\left(-\frac{\sqrt{m\triangle E/2a}}{\hbar}\right)\right]^{-1}.
\end{equation}

\subsection{non-even trigger}

The heart of the quantum measurement is the Born rule. We now study
the frequency distribution of the many repetition of measurements
for the same initial states. We would like to find a relation between
this frequency distribution and the probability read from initial
wave function of the system.

\begin{figure}
\includegraphics[width=10cm]{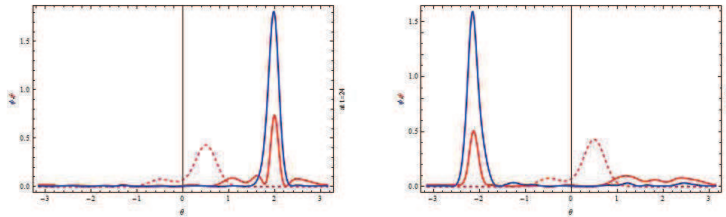}\caption{Same as Fig.\ref{fig:3}, but the system wave function $\psi_{0}(t,\theta)$
initially has unevenly superposed two Gaussian distributions with
peaks at $\theta=+1/2$ and $\theta=-1/2$.\label{fig:5}}
\end{figure}

\begin{figure}
\includegraphics[width=5cm]{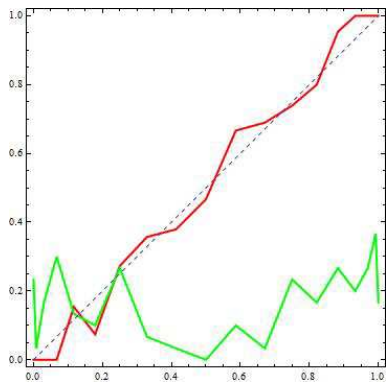}\caption{Red line: The faithfulness of the quantum measurement apparatus. The
vertical coordinate is the frequency fraction of the occurrence of
the system to be found in the negative region with the meter readout
as positive. The horizontal coordinate is $\left(\sin\alpha\right)^{2}$
of Eq.(\ref{eq:9}). Green line: The error of the quantum measurement
apparatus.The vertical coordinate is the error fraction among the
whole run of the measurements. \label{fig:6}}
\end{figure}

We set the initial wave function of the system as the general superposition
of the two Gaussian forms, $0\leq\alpha\leq\pi/2$, 
\begin{equation}
\psi_{0}(t,\theta_{0})=\sin\left(\alpha\right)\exp\left(\frac{\theta_{0}+0.5}{\triangle\theta}\right)e^{ip_{0}\theta_{0}}+\cos\left(\alpha\right)\exp\left(\frac{\theta_{0}-0.5}{\triangle\theta}\right)e^{-ip_{0}\theta_{0}}.\label{eq:9}
\end{equation}
A typical one shot measurement is shown in Fig.\ref{fig:5}. Then
we perform numerical calculations which mimic many repetition of such
measurements. A typical result is shown in Fig.\ref{fig:6}. The graph
shows the relative frequency distribution of the many repetition of
measurements, with firm correlation between the system state and the
meter readout, for the same initial states against the absolute square
of the coefficient $\left(\sin\alpha\right)^{2}$. The linear dashed
line in the figure represents the Born rule of quantum mechanics.
The green solid line shows the relative frequency that the apparatus
failed to yields the consistent result between the system and the
meter readout. The graph shows that our apparatus actually performed
quantum measurement with finite accuracy.

We have not very fine-tune the parameters to reproduce the straight
line. However generally such tuning or the calibration of the apparatus
is necessary for the efficient measurements. Precisely speaking, the
balance of the randomness $\sigma$ and the strength of trigger $\lambda$
is essential. If the former dominates the latter, the correlation
yields more flat $N$-shape, while in the opposite case yields more
vertical $S$-shape in the graph.

\section{summary and discussions}

We have discussed the synchronization and clustering of many quantum
degrees of freedom and they yield the order variable which has very
different time scale from the individual degrees of freedom. Furthermore
this order variable is localized dynamical degrees of freedom which
does not disperse. Thus it behaves more classically.

We could construct a quantum measuring apparatus utilizing this order
variable which couples to the quantum measured system. The wide separation
of time scales of them is the essence of this model. Although we could
derive very approximate Born rule, further improvements and considerations
are necessary. Some of them are as follows. 
\begin{enumerate}
\item All the calculation was based on the wave function and the Schr�dinger
evolution, which does not introduce any dissipativity which is considered
to be essential for the measurement process. Dissipative and probabilistic
nature was only prepared as the initial condition which has random
distribution of the wave functions. However this treatment is not
complete and we need the description based on the density matrix with
dissipation and fluctuations. 
\item We utilize time-dependent Hartree-Fock approximation (TDHF) since
the actual calculation would be impossible otherwise. However this
kind of self-consistent method may neglect any relevant correlations
which would be essential for the measurement process. We may need
to consider seriously the classical behavior of quantum variables\cite{key-12}. 
\item The balance of the randomness $\sigma$ and the strength of trigger
$\lambda$ was essential in our apparatus. This seems to be a general
feature in quantum measurement\cite{key-10}. 
\item There is a formalism of quantum measurement which slightly modify
the Schr�dinger equation including stochasticity and non-linearity
at the fundamental level\cite{key-13}. The basic line of thought
to derive the Born rule is the same as ours but in our case we attribute
the stochasticity and non-linearity to the detector. 
\end{enumerate}
These points should be further studied.

An essential feature of our model is the separation of time scales
of the order variable and the system state. Actually the classical
version of our model can describe the long-term dynamics of the geomagnetism\cite{key-13}.
In this version the steady global dipole mode with time scale about
million years coexists with rapid and local mode with time scale about
thousand years. The former corresponds to the apparatus readout and
the latter quantum state measured in our present model. \medskip{}

MM thank Akio Hosoya, Toshiaki Tanaka and Peter Pickl for valuable
discussions and important suggestions.


\begin{thebibliography}{1}
\bibitem{key-1} A. Pikovsky, M. Rosenblum, and J. Kurths, \textit{Synchronization:
A Universal Concept in Nonlinear Sciences }(Cambridge Nonlinear Science
Series), Cambridge University Press, Cambridge (2003).

\bibitem{key-2}G. L. Sewell, \textit{Quantum Theory of Collective
Phenomena }(Monographs on the Physics and Chemistry of Materials),
Oxford University Press, USA (1990).

\bibitem{key-13}A. Nakamichi, H. Mouri, D. Schmitt, A. Ferriz-Mas,
J. Wicht, and M. Morikawa, Oxford Journals Mathematics \& Physical
Sciences MNRAS \textbf{423}, 2977 (2012)

\bibitem{key-3}A. Peres, \textit{Quantum Theory: Concepts and Methods}
(Fundamental Theories of Physics), Springer (1995).

\bibitem{key-4}C. J. Isham, \textit{Lectures on Quantum Theory: Mathematical
and Structural Foundations}, World Scientific Pub Co Inc (1995).

\bibitem{key-10}M. Morikawa and A. Nakamichi, Progress of Theoretical
Physics \textbf{116} 679 (2006).

\bibitem{key-11}Masahiro Morikawa, preprint arxiv.org/abs/1211.1739
(2012).

\bibitem{key-12}P. Pickl, Lett. Math. Phys. \textbf{97} 151 (2011).

\bibitem{key-14}A. Bassi, K. Lochan, S. Satin, et.al., Reviers of
Modern Phys. \textbf{85} 471 (2013). \end{thebibliography}
\end{document}